\documentclass[review]{elsarticle}

\usepackage{lineno,hyperref}
\modulolinenumbers[5]

\journal{Journal of \LaTeX\ Templates}

\begin{document}

\begin{frontmatter}

\title{The Quantum Transport of  Pyrene and its Silicon-Doped Variant: A DFT-NEGF Approach}

\author{Alireza Rastkar Ebrahimzadeh\footnote{ e-mail: a\_rastkar@azaruniv.edu} }
%\ead{a_rastkar@azaruniv.edu}
\author{Badie Ghavami \footnote{ e-mail: badie.ghavami@azaruniv.edu}}
%\email{badie.ghavami@azaruniv.edu}
%\address{Molecular Simulation Lab. Faculty of Basic Sciences, Azarbaijan Shahid Madani University}
\author{\\Jaber Jahanbin Sardroodi}
\author{Sadegh Afshari}
\author{Mina Yaghoobi Notash}
\address{}
\address{Molecular Simulation Lab. Faculty of Basic Sciences, Azarbaijan Shahid Madani University}
\address{ P.O.Box : 53714-161, Tabriz, Iran}
\begin{abstract}
The quantum conductance properties of pyrene molecule and its silicone-doped variant between semi-infinite aluminum nano-chains  have been investigated by using the density functional theory (DFT) combined with the non-equilibrium Green function (NEGF) method. Electronic transport computations have been carried out in the bias voltage range valued from 0.0 to +2.0 V divided by 0.1 V step-sized intervals and under the gate potentials including -3.0, 0.0 and +3.0 V. The current-bias curves at the considered bias and gates potential show regions with negative differential resistance (NDR). The  effects of the variations of the gates on the NDR characteristics including the number of NDR peaks, bias range and current maxima's at the peak have been discussed and the potential applicability of the devices as nano-switches and multi-nanoswitches have been discussed. The transmission spectrum along with the density of states (DOS) and projected DOS (PDOS) have also been presented and transmission variations has been discussed in terms of the DOS and PDOS variations.Quantum conductance at zero bias versus gate potential has been also presented and discussed.
\end{abstract}


\end{frontmatter}

\linenumbers

\section{Introduction}
In recent years, charge and electron transport via molecules put among electrodes has been attracted increasing attention for basic reasons and for because it may forms fundamental nano and molecular electronics devices \cite{nitzan, heath, tour, joachim, Darancet}. The capability to calculate the electronic structure of including molecules by using quantum mechanics and also the transport properties of these devices is momentous and useful in condensed matter physics, electronics and other related fields of sciences and applications.
Among the powerful and efficient methods for the investigation of the electronic transport properties is the combination of density functional theory (DFT) and the Non equilibrium Green's function (NEGF) methods. Mainly, this method was extended to self-consistently computing ballistic conduction including inelastic scattering \cite{Paulsson} occurring in nanoscale devices such as molecular and atomic semiconductors (FET's, MOSFET's molecular rectifiers and so on) under bias voltage and at various gate potentials \cite{Haug, Caroli, Schwinger, Keldysh, Reddy,Wang, Rocha}.
Recently, some researchers have focused on small aromatic polycyclic hydrocarbons as highly conductive molecular devices \cite{billic}.\\
%%%%%%%%%%%%%%%%%%%%%%%%%%%%%%%%%%%%%%%%%%%%%%%%%%%
In this research work, the transport properties including current-bias values and transmission spectrum of the pyrene molecule and its silicon substituted derivative lied between aluminum semi-infinite atomic chains have been computed. The computations have been carried out in the bias voltage range of 0.0 to 2.0 V divided by 0.1 V intervals and under the gate potential values of -3.0, 0.0 and +3.0 V by the help of DFT-NEGF method. The results show that there are negative differential resistance(NDR)\cite{mahmoud} regions in the computed current-bias voltage(I-$V_b$) curve of the considered systems. The transmission spectrum along with the computed electronic properties such as the Density of States (DOS) have been used for interpretation and explanation of the observed I-$V_b$ including NDR behavior of the studied systems.
The results of the present computations show that the NDR characteristics including the number of peaks, the bias range of NDR region and maximum current amplitude would be varied by changing the gate potential. The gate controllable nature of a device is a useful property in its applicability. The results suggest that the studied systems would be used as nano-switch and multi-nanoswitch in the consider bias voltages and gate potentials.\\
%%%%%%%%%%%%%%%%%%%%%%%%%%%%%%%%%%%%

\section{Theory and Calculation Method}
In order to computational investigation of the quantum and electronic transport properties of the systems including molecular junctions and scattering regions one have to solve the non-equilibrium Green's functions (NEGF) equations using the wave functions obtained by the density functional theory (DFT) \cite{kondo, ozaki}. It briefly is explained in the following paragraphs.\\
In the DFT based solving of the NEGF equations, the system under consideration is divided into three part composed from two semi-infinite left and right electrodes,and a scattering region, denoted by L, R and S parts, respectively. The S-part has been composed from the considered molecule as central region and the atomic layers of the left and right electrodes. The Hamiltonian matrix $H$ and the overlap matrix $S_{overlap}$ for this device are given by:
\begin{equation}
\mathbf{H}=\left(
\begin{array}{ccc}
H_{L} & H_{LS} & 0 \\
H_{SL} & H_{S} & H_{SR} \\
0 & H_{RS} & H_{R} \\
\end{array}\right),
\hspace{0.3cm}
\mathbf{S_{overlap}}=\left(
\begin{array}{ccc}
S_{L} & S_{LS} & 0 \\
S_{SL} & S_{S} & S_{SR} \\
0 & S_{RS} & S_{R} \\
\end{array}\right).
\end{equation}
where matrix elements of the Hamiltonian, $H_{S}$, is of size $N\times N$, in which $N$ is the number of basis functions in the scattering region calculated under within the desired boundary conditions imposed by the electrodes. In this framework, it is also assumed that there is no interactions between the L and the R parts.\\
Retarded Green' function, $G^r$, for the S-part and the retarded self-energy for the L and R parts are defined as
$G^r(E)=lim_{\eta\rightarrow 0^+}(\varepsilon+i\eta-H)^{-1}$ using the following equations:
\begin{equation}
G^r_{S}(E)=(ES_S-H_{S}-\Sigma ^r_{L}(E)-\Sigma ^r_{R}(E))^{-1}
\end{equation}
\begin{equation}
\Sigma ^r_{L}(E)=(ES_{SL}-H_{SL})G^r_{L}(ES_{LS}-H_{LS})
\end{equation}
\begin{equation}
\Sigma ^r_{R}(E)=(ES_{SR}-H_{SR})G^r_{R}(ES_{RS}-H_{RS})
\end{equation}
where $E=\varepsilon+i\eta$ (that $\eta\rightarrow 0^+$) and $G^r_{L(R)}(E)=(ES_{L(R)}-H_{L(R)})^{-1}$  are the retarded surface Green's function for the L and R parts. The density matrix\cite{Chen} is computed by the help of the equation (2-4).
\begin{equation}
\rho_{ij}=-\frac{i}{2\pi}\int_{-\infty}^{+\infty}dE G_{S;ij}^<(E)
\end{equation}
in which $G^<(E)$is the lesser Green's function defined as $G^<(E)=G^r(E)\Sigma ^<(E)G^a(E)$, where $G^a(E)$ is the advanced Green's function and $\Sigma ^<(E)$ is lesser self energy. It is also defined that $G^{a\dagger}(E)=G^r(E)=[ES-H+\Sigma^r(E)]^{-1}$ where $\Sigma^r$ is the retarded self energy.
In the equilibrium case in which bias voltage is zero, the density matrix is calculated by\cite{ozaki}
\begin{equation}
\rho_{ij}=-\frac{1}{\pi}\int_{-\infty}^{\mu}dE G_{S;ij}^r(E)
\end{equation}
where $\mu$ is the chemical potential. The charge density is calculated in the scattering region in the presence of left and right electrodes as the followings
\begin{equation}
\rho=\sum_{ij}\phi_i(\textbf{r})\rho_{ij}\phi^{\ast}_{j}(\textbf{r})
\end{equation}
where $\phi_i(\textbf{r})$ is the numerical pseudo-atomic orbital.
The Hamiltonian $H_{S}$ is formed from the kinetic energy term and potential term composed from the nonlocal part of the pseudo potential and other potential terms given by the following equation
\begin{equation}
V(\textbf{r})=V_H(\textbf{r})+V_{xc}(\textbf{r})+V_{loc}(\textbf{r})
\end{equation}
where $V_{xc}(\textbf{r})$ and $V_{loc}(\textbf{r})$ are the exchange-correlation potential and the local part of the atomic pseudopotential respectively and $V_H(\textbf{r})$ is Hartree potential obtained by solving the Poisson equation $\nabla^2 V_H(\textbf{r})=-4\pi\rho(\textbf{r})$.\\
This procedure is iterated until self consistency is achieved. Then, the transmission is given by
\begin{equation}
T(E)=Tr[\Gamma_{L}(E)G_{S}^r(E)\Gamma_{R}(E)G^{a}_{S}(E)]
\end{equation}
and
\begin{equation}
\Gamma_{L(R)}(E)=i[\Sigma_{L(R)}^{r}(E)-\Sigma_{L(R)}^{a}(E)]
\end{equation}
where $\Sigma_{L(R)}^{a}$ are the advanced self-energy for the left and right region. The current through the atomic scale system can be calculated
from the corresponding Green's function and self-energies via the Landauer formula,\cite{Landauer, Buttiker}
\begin{equation}
I(V)=\frac{2e}{h}\int_{-\infty}^{\infty}T(E,V_b)[f^L(E-\mu_L)-f^R(E-\mu_R)]dE
\end{equation}
where $\mu_L$ and $\mu_R$ are the electrochemical potentials of the left
and right electrodes respectively, i.e.($\mu_L-\mu_R=eV_b$) and $f^{L(R)}(E-\mu_{L(R)})$ are the corresponding electron distribution of the two
electrodes, and $T(E,V_b)$ is the transmission
coefficient at energy $E$ under bias voltage $V$.\\
The transmission coefficient $T(E)$ can be decomposed into the contribution of $n$ eigenchannels,
\begin{equation}
T(E)=\sum_{n}T_{n}(E)
\end{equation}
For the system in equilibrium state, conductance can be obtained by the transmission coefficient $T(E)$ at the Fermi level $E_f$,
\begin{equation}
G=G_0T(E_f)=G_0\sum T_{n}
\end{equation}
where $G_0=\frac{2e^2}{h}$ is the conductance quantum ($7.748091733\times 10^{-5} S$)\cite{Taylor,Brandbyge}. \\
In this paper all of the calculations have been performed within fully self-consistent non-equilibrium Green's functions (NEGF) and density
functional theory(DFT) \cite{datta1, datta2} using Open Source package for Material eXplorer (OPENMX-3.6) software package \cite{ozaki, openmx}. The Perdew-Burke-Ernzerhof parametrization of the generalized gradient approximation(GGA-PBE) to exchange-correlation functional are used \cite{martin, Perdew}. The electronic temperature is to 300 kelvin and the spin polarized
calculation is performed. In this research the cutoff energy is set to 120 (Ryd) and the Fermi level $E_f$ is set to zero. The cutoff radius $r_c$ of the pseudo-atomic orbitals used in the calculations is presented as the Table 1.\\
\begin{table}[ht]
\caption{The cutoff radius $r_c$ of the two systems.}
\vspace{0.5cm}
\centerline{
\begin{tabular}{|l|c|c|c|c|}
\hline\hline
     Atom  & AL    & C  & Si  & H \\
\hline
$r_c(au)$ & 7.0     & 7.0   &7.0  & 5.0 \\
\hline\hline
\end{tabular}
}
\end{table}
Figure 1-A and B have been contained the considered devices composed from the scattering region and left and right electrodes. These figures show that the pyrene molecule and its silicon-doped variant form the central region of system A and B, respectively. These devices were also composed from semi-infinite aluminum atomic chains as L and R electrodes. The distance between Al atoms of electrodes and carbon atoms of central region is obtained by the energy optimization of the considered devices. Aluminum atomic chain is directed along the z axis.
\begin{figure}
	\centering
	    \includegraphics[angle=0, width=0.8\textwidth]{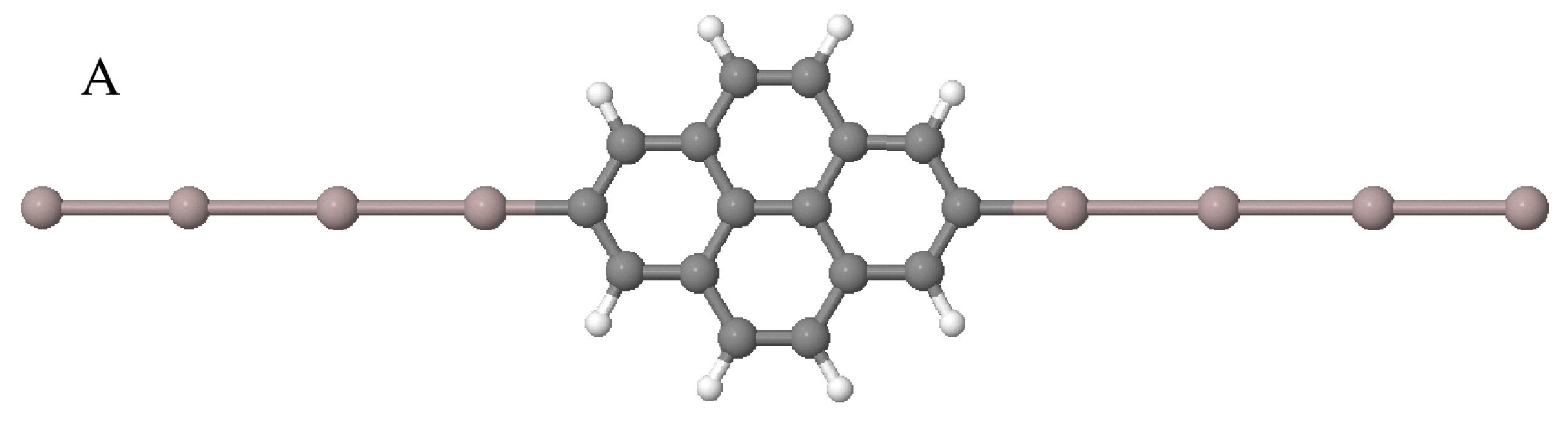}
		\includegraphics[angle=0, width=0.8\textwidth]{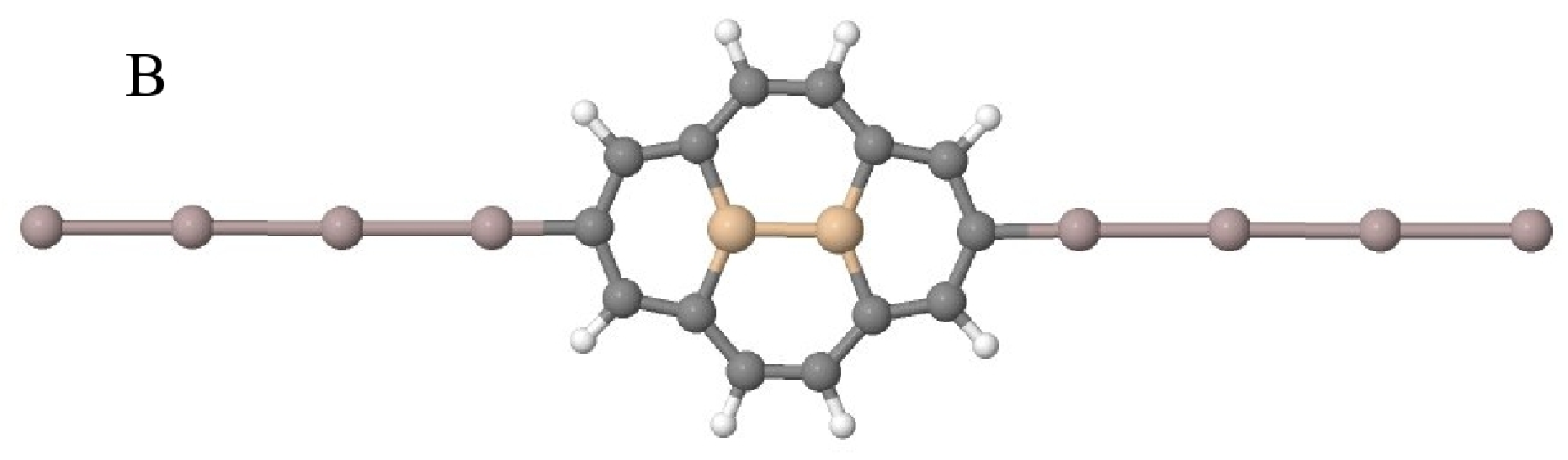}
        \caption{A. device with pyrene molecule as central region, B. device with silicon-doped pyrene variant}
\label{quantum conductance}
\end{figure}
The calculations have been carried on in the bias voltage range of 0 to 2.0 v with 0.1 V step-sized intervals applied in order to shifting the chemical potential of the electrodes and under the gate potentials including -3.0, 0.0 and +3.0 V applied on the scattering region.
%%%%%%%%%%%%%%%%%%%%%%%%%%%%%%%%%%%%%%%%%%%%%%%%%%%%%%%%%%%%%%%%%%%%%%%%%%%%%%
\section{Results and discussion}
\label{3}
The self-consistently calculated currents versus bias voltage (I-$V_b$) for the studied
systems at the considered gate voltages lied at the range of $- 3.0, 0.0$
and  $3.0$ V have been presented as Figure 2. Figures 2-a, 2-b and 2-c have been contained the I-$V_b$ for the studied systems in gate potentials 0.0, 3.0 and -3.0 V, respectively.
\begin{figure}
	\centering
              \includegraphics[angle=270.0, width=1.1\textwidth]{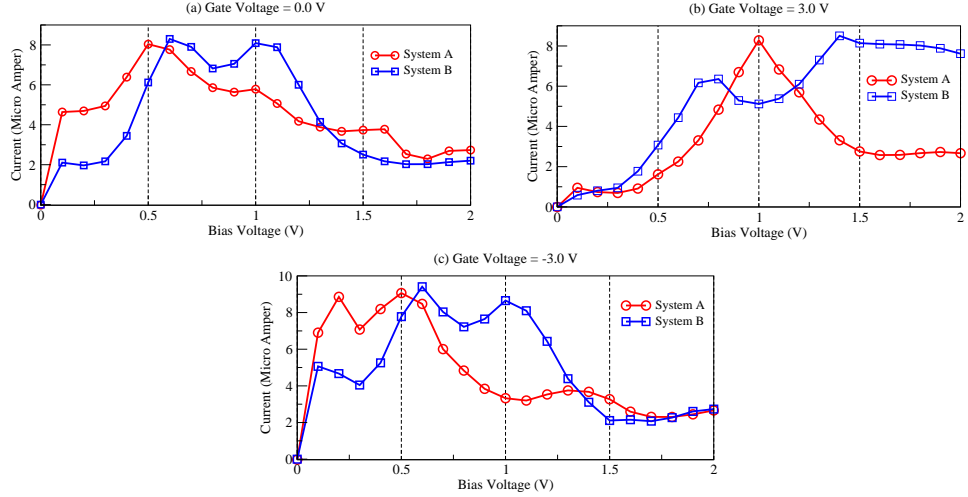}
              \caption{Computed current at the bias voltages 0 to 2 V for the considered devices; a. $V_{gate}$ = 0.0, b.  $V_{gate}$ = +3.0 and c.  $V_{gate}$ = -3.0 $V$}
\label{2}
\end{figure}
 These figures show
that the I-$V_b$ behavior would be varied by substitution carbon atoms
with silicon atoms and also with variations of the gate voltage. The
variations are to generate or shifting the Negative Differential Resistance
(NDR) regions. The results presented in these figures also show that the initial trend of the I-$V_b$ curves for the
studied systems are similar. This initial trend consists of a nearly linear
increasing in current with the bias leading to a non-linear behavior with a
NDR. The variations resulted by substitution of carbon atoms and/or variation
of the gate voltage appear beyond of first NDR. These variations are included
a linear decrease of the current with bias in system A at $V_{gate}$=0.0 and 3.0 V;
and a nearly constant current for system B at $V_{gate}$=3.0 V. These figures also show that, it was generated one or tow
additional NDR regions at the remaining cases including system A under the gate of -3.0 and system B under gates values of 0.0 and -3.0 V.
The generation of additional one and tow NDR regions beyond the
observed one, is not reported in the literature, yet. \\
The double NDR observed for system B at $V_{gate}$=0.0 and -3.0 and for system A at
$V_{gate}$=-3.0 suggests that these devices could be used as multi-switches at the
corresponding bias ranges. The appearing of the observed first or second NDR's
is due to the variations of the gate voltages and this is a great advantages
for these devices. This give us the ability of changing the behavior of the
device without changing its structure. Changing the value of a
gate potential on a device is a simpler task in comparison to the changing its
structure.\\
The transmission spectrum and the density of states, $(DOS)$, in zero bias and zero gate for System A and System B have been represented in Figures 3 and 4, respectively.
\begin{figure}
	\centering
		\includegraphics[angle=270, width=0.9\textwidth]{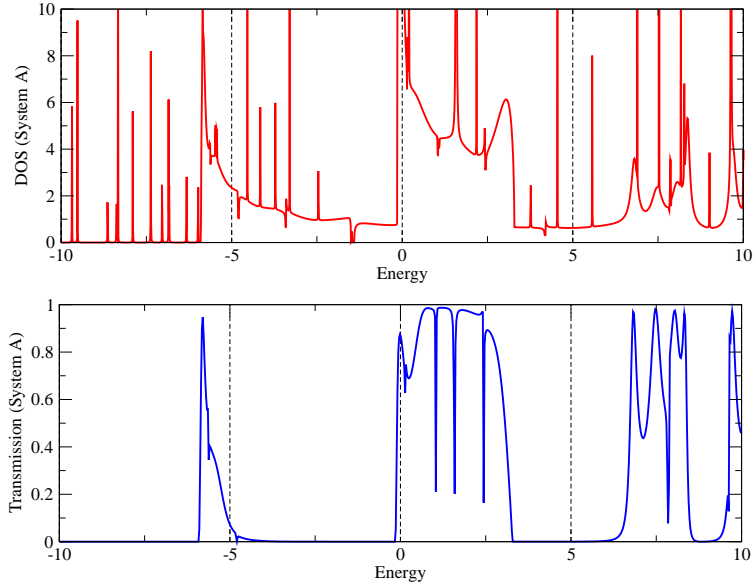}
		\caption{Density of states (DOS) and Transmission(T) spectrum of the system A, at zero bias and gate}
\label{3}
\end{figure}
\begin{figure}
	\centering
		\includegraphics[angle=270, width=0.9\textwidth]{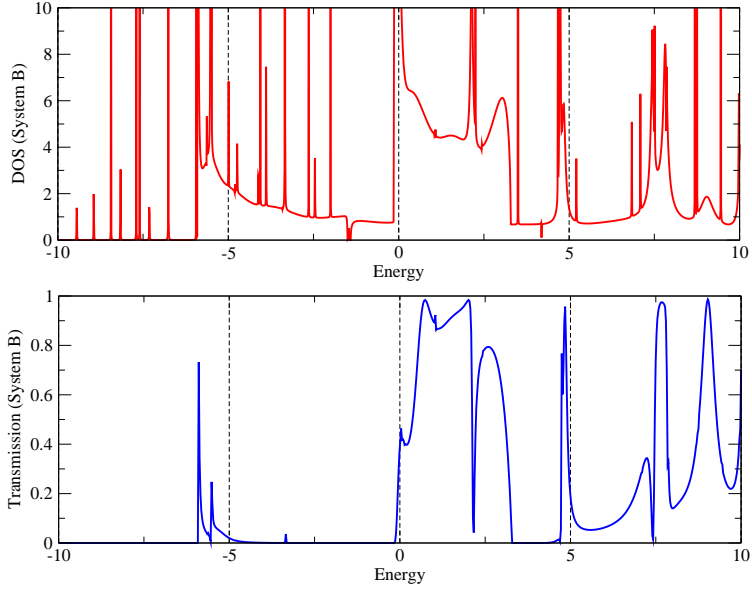}
		 \caption{Density of states (DOS) and Transmission(T) spectrum of the system B, at zero bias and gate}
\label{4}
\end{figure}
%%%%%%%%%%%%%%%%
These figures show that the transmission spectrum and the $DOS$ spectrum are in a good agreement with each other.
The transmission peaks below the Fermi level, $(E\leq E_f)$, can be demonstrated by the occupied molecular orbitals and those above the Fermi level, $(E\geq E_f)$, by the unoccupied molecular orbitals. These figures also show that the Fermi level lies at the vicinity of the LUMO, so the bias in which the current raised, is correspond to the energy transporting the electrons through the LUMO. This happens due to the effective overlap of the states corresponding to the first LUMO of the scattering molecule and the energy states of the
 electrodes.\\
Figures 3 and 4 shows Fano and Wigner resonance position in the transmission spectrum\cite{fano,wigner}. These resonances refer to the very sharp variations in the transmission curves. The so called Fano resonance is defined as the vanishing of the values of the transmission function rapidly.\\
The Fano-type resonance in the molecular systems is an effective property that enables us to design the molecular FET with a relatively high on-off ratio\cite{sowa}. Furthermore, the results also show that the considered devices would be used for switching aims in design the Nano-electronics circuits.\\
%%&&&&&&&&&&&&&&&&&&&&&&&&&&&&&&&&&&&&&&&&&&
Figures 5-a and 5-b contain the transmission spectrum and projected DOS (PDOS) for the system A and B respectively. 
\begin{figure}
	\centering
		\includegraphics[angle=270, width=0.9\textwidth]{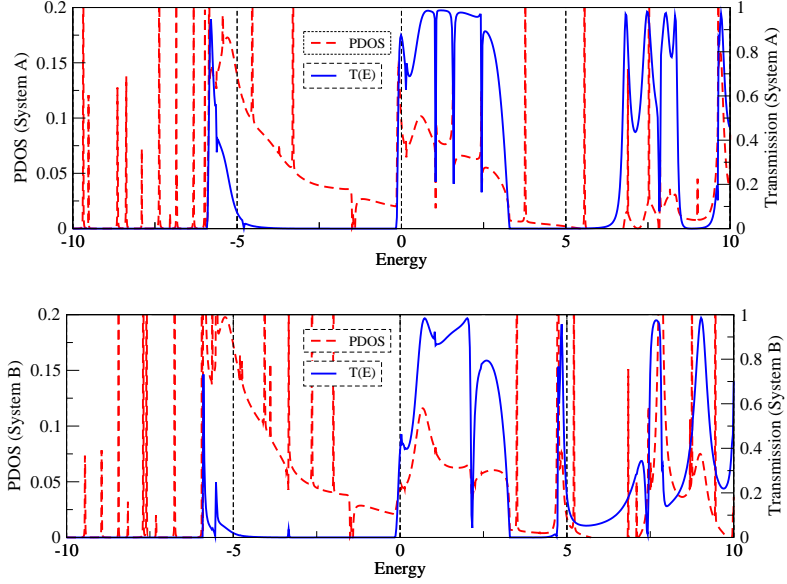}
\caption{projected density of states (PDOS) and transmission spectrum
at zero bias and gate; a. System A, b. system B}
\label{quantum conductance}
\end{figure}
%%%%%%%%%%%%%%%%%%%%%%
\begin{figure}
	\centering
		\includegraphics[angle=270, width=0.9\textwidth]{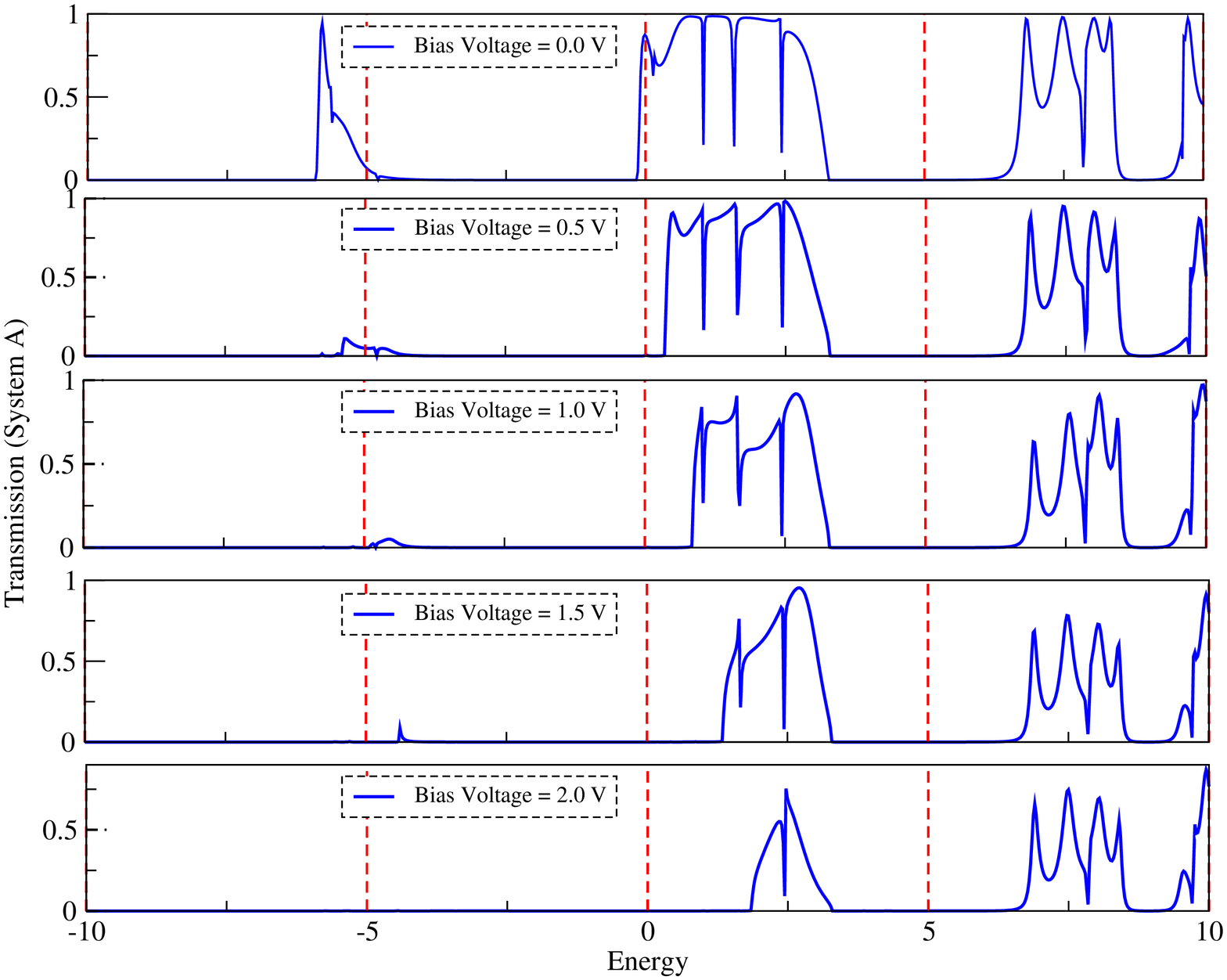}
		\includegraphics[angle=270, width=0.9\textwidth]{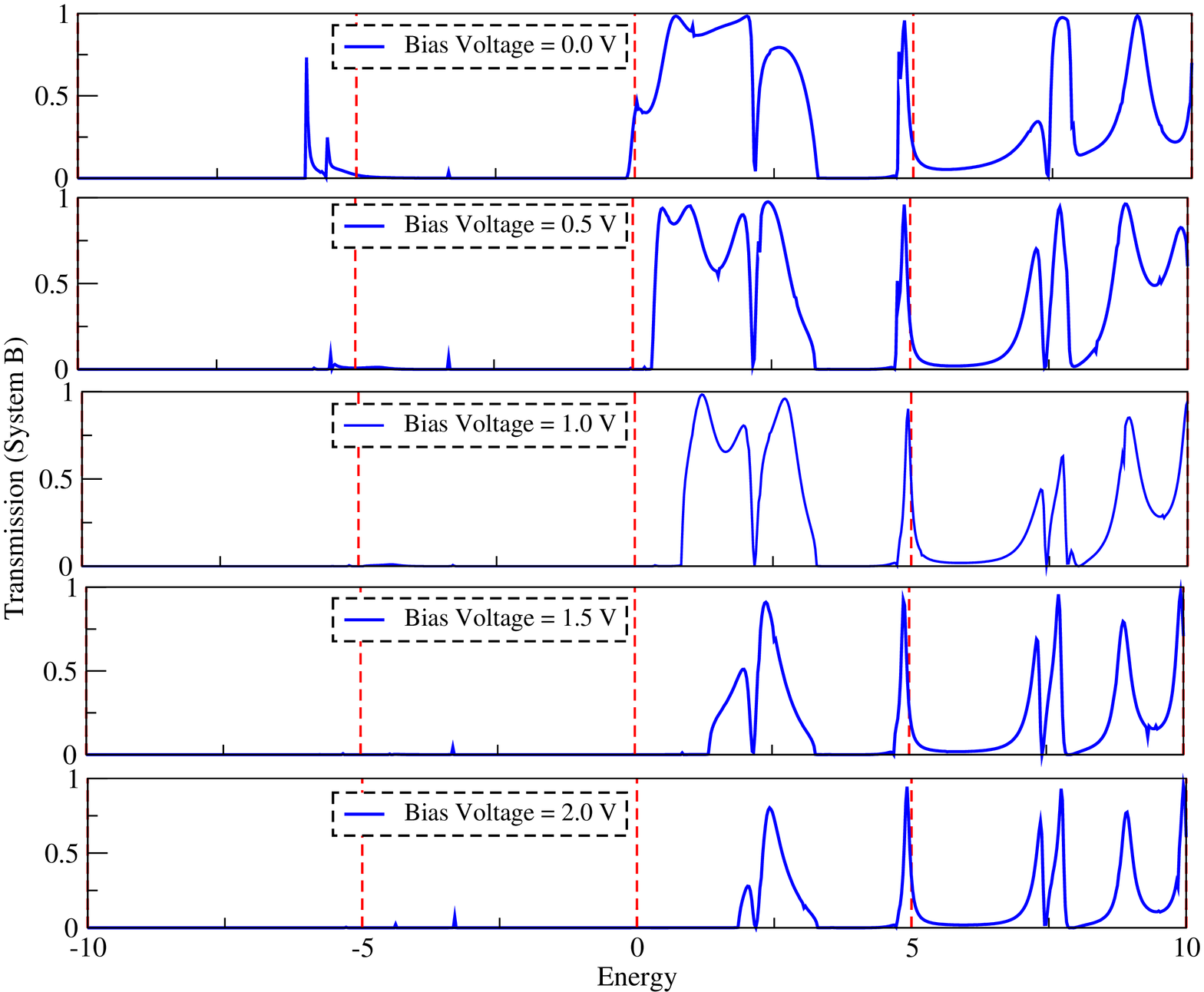}
                  \caption{Transmission spectrum of systems A and B at the bias voltages including  0.0, 0.5 1.0 1.5 and 2.0 V at zero gate voltage}
\label{quantum conductance}
\end{figure}
This quantity is defined as\cite{san}
\begin{equation}
N_m=-\frac{1}{\pi}Im(Tr_m[\textbf{G}_{cc}(E+i\eta).\textbf{S}_{cc}])
\end{equation}
where $\textbf{G}_{cc}$ and $\textbf{S}_{cc}$ is Green's Function for scattering region and overlap matrix, respectively. Also $Tr_m$ means the trace is performed only on the molecular part of matrix.\\
The PDOS will give us information on how much the basis orbitals in the molecule contribute to the eigenstate of the whole open system and how strongly the molecule couples with the electrodes at a certain energy.\\
These figures show that, the transmission and the PDOS are in good agreement with each other. However, there is not good agreement between transmission and PDOS in few energies the localized states of central region and electrodes were not coupled in them. The transmission peak at the spectrum of Figure 5 reveals strong coupling between the localized states of the central region and electrodes.\\
The self-consistently calculated transmissions for the studied systems at zero gate voltage in the considered biases including 0.0, 0.5, 1.00, 1.50 and 2.00 V have been presented as Figure 6.  Applying a bias voltage, Vb, on the electrodes shifting transmission values corresponding to the Fermi energy according to the following equation obtained by San-Huang et al. \cite{san}:
\begin{equation}
T(E,V_b)=Tr[\Gamma_L(E+\frac{ev_b}{2})G_{cc}(E)\Gamma_R(E-\frac{ev_b}{2})G^{\dagger}_{cc}(E)]
\end{equation}
Figures 6 show that increasing bias voltage, decreasing the Fano resonance and unresonance energies. These figures also show that the value of energy shift for fano resonance and unresonance is considerable for energies around Fermi level. \\
%%%%%%%%%%%%%%%%%%%%%%
\begin{figure}
	\centering
		\includegraphics[angle=270, width=0.9\textwidth]{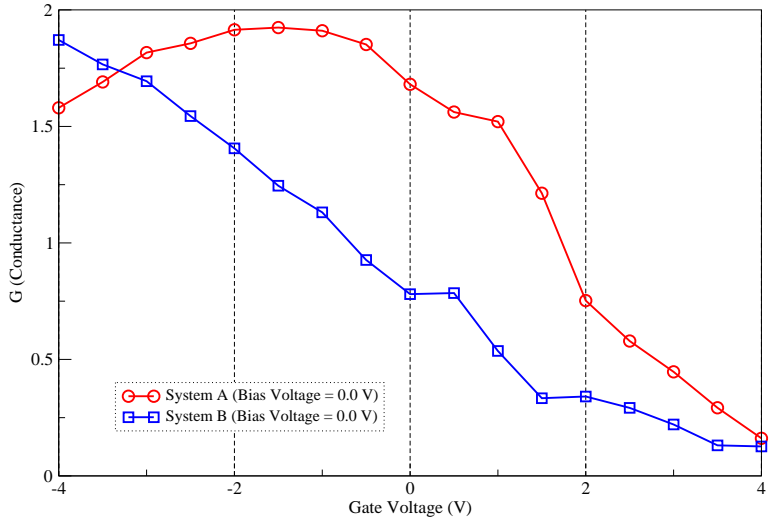}
                  \caption{ The variation of quantum conductance, $G$, of the systems A and B with the gate voltage, $V_g$}
\label{quantum conductance}
\end{figure}
%%%%%%%%%%%%%%%%%%%%%%%%%%%
Figure 7 have been presented the computed conductance for the considered systems with respect to the applied gate potential at the zero bias. This figure show that the conductance varies non-linearly with gate potential for system A and near linear for system B. The results of this figure show also that the numerical values of the conductance of system A are larger than the conductance of system B and the differences between tow conductances are vanished when gate goes larger.
%%%%%%%%%%%%%%%%%%%%%%%%%
\section{Conclusion and remarks}
\label{4}
DFT+NEGF method has been employed for computational study of electron transport of devices composed from aluminum nano-chain electrodes and pyrene and silicon-doped pyrene as scattering region in the bias range of 0 to +2 eV and under the gate potentials covering -3.0, 0.0 and +3.0 eV. The current-bias properties have been discussed in terms of transmission spectrum, DOS and PDOS. The negative differential resistance properties of the studied systems have been treated on at the considered biases and their variations with respect to applied gates have been discussed. The applicability of the devices as nano-switches and multi-nanoswitches have also been discussed on the basis of the current-bias curves.The variations in transmission values interpreted in terms of localized states couplings using the PDOS and DOS spectrums. The conductance of the studied systems with respect to applied gates at zero bias have been presented and discussed.\\
%%%%%%%%%%%%%%%%%%%%
\section{Acknowledge}
\label{5}
This research was supported by a research fund "$No: 217/D/5666$"
from Azarbiajan Shahid Madani university.\\

\end{document}